# Imaging-free object recognition enabled by optical coherence

Yixuan Tan[1†], Xin Lei[2†], Ken Xingze Wang[2†*], Shanhui Fan[3], and Zongfu Yu[1*]

[1]University of Wisconsin, Department of Electrical and Computer Engineering, Madison WI, 53706; [2]Coherent AI LLC, Redwood City, CA 94065; [3]Stanford University, Ginzton Laboratory, Department of Electrical Engineering, Stanford, CA 94305

† These authors contribute equally to the work

K.X.W. ken@coherent.ai

Z.Y. zyu54@wisc.edu

**Abstract:** Visual object recognition is one of the most important perception functions for a wide range of intelligent machines. A conventional recognition process begins with forming a clear optical image of the object, followed by its computer analysis. In contrast, it is possible to carry out recognition without imaging by using coherent illumination and directly analyzing the optical interference pattern of the scattered light as captured by an image sensor. Here we show that such direct visual recognition can overcome traditional limitations of imaging optics to realize excellent recognition without focusing, beyond diffraction limit, or in the absence of direct line-of-sight.

# Main

Optical object recognition is of importance in a wide range of applications such as face recognition, industrial inspection, and autonomous driving. In standard optical object recognition, one starts with forming an optical image of the object, followed by computer analysis of the resulting image. However, an image almost always contains much more information than that required for recognition purposes. Thus, from a basic information theory perspective, for recognition purposes, there is no fundamental reason that one needs to start with a well-formed optical image.

Motivated by the above argument, in this study, we demonstrate an optical object recognition system that eliminates the need for imaging. A simple schematic of our setup is shown in Fig. 1, where a set of objects, a set of hand-written digits in our case, is illuminated with a coherent laser beam. The light scattered from the objects then forms speckle patterns, which we detect using an image sensor. These speckle patterns certainly do not look like the hand-written digits. Nevertheless, we show that a neural network trained directly on the speckle patterns is sufficient to recognize the digits. Such recognition, moreover, can be achieved even when imaging is difficult. As illustrations, we show successful recognitions beyond the diffraction limit and in situations where there is no direct line-of-sight between the objects and the image sensor.

Related to our study, there is considerable literature showing that useful information can be retrieved from speckle patterns [1-20]. Most of these studies sent an image through a scattering medium and recovered the image by analyzing the resulting speckle patterns or holograms. For recognition purposes here, however, image recovery is not necessary, as we demonstrate in this study. Recognitions that are based directly on speckle patterns have been considered in [21-25]. Here we show a systematic study of object recognition under conditions that are challenging for imaging optics.

Imaging-free visual recognition leverages upon the optical interference enabled by coherent illumination. As shown by the simulations in Fig. 1b, if an image sensor were placed at a certain distance away from the object illuminated by coherent light, the measured signals would exhibit rich features created by optical interference. These features, commonly known as speckle patterns, evolve as the distance changes. The information carried by these patterns is not obvious to a human, but after training (see Supplementary Section 3), a neural network can recognize the digits with over 90% accuracy at all locations. This interference is absent under incoherent illumination such as lamps. The field intensity is, in general, nearly uniform and contains no useful information, considering the relative sizes of image sensors and objects. Fig. 1c shows the calculated distribution of light intensity with 16-bit accuracy. We train various neural networks. As expected, none of them can recognize the object. The recognition accuracy on the test set is around 10%, which is equal to the accuracy of a random guess.

An imaging system provides point-to-point mapping from the object to the image sensor. Without an imaging system, direct light detection is equivalent to multiplex sampling of the object. As shown in Fig. 1d, a pixel at location $\vec{r}$ on the sensor chip receives light from all over the scene $I(\vec{r}) = \left| \int E_s(\vec{R}) \frac{\exp(-i\vec{k}\cdot(\vec{R}-\vec{r}))}{|\vec{R}-\vec{r}|} d\vec{R} \right|^2$, where $E_s(\vec{R})$ represents the complex amplitude of light source at location $\vec{R}$ in the scene and $\vec{k}$ is the wavevector. This multiplex sampling is generally not invertible, which means that solving for $E_s(\vec{R})$ is an under-determined problem [26-32]. However, considerable information is still retrievable because the multiplex bases $\frac{\exp(-i\vec{k}\cdot(\vec{R}-\vec{r}))}{|\vec{R}-\vec{r}|}$ are substantially uncorrelated due to rapid spatial variation introduced by the phase term $\vec{k}\cdot(\vec{R}-\vec{r})$. A trained neural network can effectively use such information to perform correct recognition. This is not the case for incoherent light, where the light intensity is given by $I(\vec{r}) = \int \left| E_s(\vec{R}) \frac{1}{|\vec{R}-\vec{r}|} \right|^2 d\vec{R}$. The multiplex bases $\frac{1}{|\vec{R}-\vec{r}|}$ lack the phase variation and are highly correlated (see discussion in Supplementary Section 1). In fact, for typical object sizes, all pixels on the image sensor obtain the same nearly uniform signal, as shown in Fig. 1c. Direct measurement of incoherent light fails to preserve most of the information in the object. To retrieve information in incoherent light, an imaging system is needed. A pixel on the image sensor measures light from one location from the scene, and point-wise sampling preserves the spatial information.

Bypassing imaging, direct visual recognition opens exciting opportunities to tackle some of the most difficult challenges resulting from fundamental limitations of imaging optics. Below, we show three demonstrations in which coherent light is used to recognize objects that are (1) much out-of-focus, (2) beyond diffraction limit, and (3) without direct line-of-sight. In all three cases, traditional imaging technique would be difficult, whereas object recognition can still be performed successfully.

## Results

**Focusing-free recognition.** An imaging system works by using a point-like point spread function (PSF). This PSF increases in its spatial extent when an object moves away from the focus. Consequently, the image of an object degrades as the object moves away from the focus. On the other hand, even in the absence of an imaging system, objects illuminated by coherent light can be recognized without the limitation of the focus.

Fig. 2a, b shows the experimental setup that contrasts with the difference between the cases of coherent and incoherent illuminations. In both cases, the object is a reflective LCD screen that displays hand-written digits one at a time. The object is illuminated by a light source, and part of the scattered light is then collected by a Quantalux sCMOS camera with 1920-by-1080 pixel resolution (Thorlabs CS2100M). We vary the distance between the object and image sensor from 1 to 8 m. In each location, we display 10,000 different hand-written digits. The resulting image, as observed by the image sensor, is then fed to a trained neural network in order to recognize the digits.

In the incoherent case, the light source is a white lamp. We place a compound lens in front of the image sensor. The focus of the lens system is fixed 1 m away from the image sensor. Thus, the object starts at the focus location and then moves away. The images collected by the sensor well-reproduce the object when it is placed at the focus, but become increasingly blurred as the object moves away from the focus, as shown in Fig. 2c. The recognition accuracy starts at over 90% when in focus and gradually drops to around 10% when going out of focus (Fig. 2f).

In the coherent case, the light source is an expanded laser beam (Thorlabs HNL100RB randomly polarized red laser at 632.8 nm wavelength). Moreover, in contrast to the incoherent case, here we do not use a lens system in front of the image sensor. Instead, the scattered light is directly collected by a CMOS image sensor, as shown in Fig. 2b. The resulting images at the sensor have no resemblance of the digits on the display, and instead, exhibit speckle patterns due to the strong interference effect. This interference pattern changes significantly as the object–sensor distance varies (Fig. 2d). Despite the lack of apparent resemblance of the images with drastically different patterns at different distances, a trained neural network can correctly recognize the digits with an accuracy of above 90%. Importantly, the same neural network works for all distances considered here, and thus, there is no need to re-train the neural network as the distance varies. The experimental results are also consistent with the simulation results, which are shown in Fig. 2e (See Supplementary Section 2 for a detailed discussion on simulation).

Comparing the coherent and the incoherent cases, we see that the capability for recognition in the out-of-focus situation is directly enabled with the use of coherent illumination. With coherent illumination, successful recognition does not require the formation of an image.

We contrast our demonstrated capability for recognition through speckle patterns with previous studies that used laser speckles for image recovery [5-7], some of which also used various neural network approaches. For image recovery, the details of the algorithms, such as the parameters of the trained neural network, depend sensitively on the operating condition [6]. For example, if the distance between the object and the sensor changes, the speckle patterns also change.

Consequently, the neural network used for image recovery typically needs to be retrained. In contrast, our results here, where we train a single neural network to perform successful recognition as the distance varies, indicate a significantly enhanced robustness in recognition as compared to image recovery.

**Recognition beyond the diffraction limit.** Standard imaging systems are subject to the diffraction limit. Resolving a feature of size $\Delta x$ requires a lens with a minimal aperture size of $D_{min} \sim \frac{\lambda L}{\Delta x}$, where $\lambda$ is the wavelength and $L$ is the distance between the object and the aperture, as shown in Fig. 3a. For an object with a minimum feature size $\Delta x$, the quality of the image degrades significantly when the aperture size of the imaging system falls below the minimal aperture size required. Therefore, a recognition system that is based on imaging should degrade significantly when the aperture size of the imaging system falls below what is required by the diffraction limit. In contrast, in our approach, the success of recognition is not strongly correlated with the image quality. Consequently, we expect that our approach should enable recognition well below the diffraction limit.

We support the above claim through both simulation and experiment. The setups shown in Fig. 3 are identical to those shown in Fig. 2, except for the addition of an aperture. For the incoherent case (Fig. 3b), which serves to illustrate a recognition system that is based on imaging, the object is placed at the focal point of the lens and the aperture is placed in front of the lens. For the coherent case (Fig. 3a), we choose the same distance between the object and the image sensor as that in the incoherent case. The aperture is placed in front of the image sensor. We assume the minimum feature size $\Delta x$ to be 1/10 of the object size. The aperture size is varied so that the system can operate either above or below the diffraction limit associated with such minimum feature size.

Fig. 3c shows the simulation results for both the incoherent and coherent cases. We use Fourier optics techniques to compute the propagation of light through space and aperture. For the incoherent case (setup shown in Fig. 3b), a high recognition accuracy of above 90% is achieved for a large aperture, for which a high-quality image can be formed. The recognition accuracy, however, drops sharply when the aperture size falls below the minimum aperture size, as required by the diffraction limit. For the coherent case (setup shown in Fig. 3a), high accuracy is retained even when the aperture is well below the minimum aperture size $D_{min}$, as required by the diffraction limit (Fig. 3c). The recognition only fails when the aperture size is more than 100 times smaller than that required by the diffraction limit.

Fig. 3d shows the experimental result for the coherent case. In agreement with the simulation, the experiment shows the recognition accuracy decreasing slowly as the aperture size decreases and remaining high even when the aperture size is 70 times smaller than what is required by the diffraction limit. In this experiment, reducing the aperture size reduces the area of the exposed regions on the image sensor. The smallest aperture used corresponds to an area of 100 μm × 100 μm or 20 × 20 pixels on the CMOS image sensor. To exclude the influence of data dimensionality in Fig. 3d, we down-sample all images to the same dimensions as those before feeding into a neural network.

**Recognition without direct line-of-sight**. Seeing around the corner is another challenging scenario for imaging optics. Imaging methods usually involve non-traditional measurement of light, for example, by measuring time-of-flight [38-39] or in a setup that preserves the memory effect [1, 3-4, 20]. Bypassing imaging, visual recognition around the corner can be accomplished straightforwardly using coherent illumination and standard CMOS image sensors.

Fig. 4a shows the experimental setup, which is the same as that in Fig. 2b, except that we place walls such that there is no longer a direct line-of-sight between the image sensor and the illuminated object. The wall scatters light diffusively, and part of the scattered light is captured by the image sensor. Fig. 4b shows the representative images taken by the CMOS image sensor for different objects. All images are full of speckles, indicating strong scattering and interference effects. Despite the lack of direct line-of-sights, Fig. 4c shows the confusion matrix of recognition results on test objects with an overall recognition accuracy of well above 80%. The major reason for lower accuracy than previous setups comes from the difficulty to achieve perfect mechanical stability of the walls in the experiment.

Because imaging is not required here, the direct recognition method around the corners is generally simpler and more robust than imaging-based approaches. For example, time-of-flight measurement [38-39] requires substantial resources such as nanosecond pulsed lasers and single photon detectors. In contrast, only standard CMOS image sensors and lasers are used here. Time-of-flight also takes much longer to measure, whereas our approach is a single-shot measurement. Compared to imaging methods that are based on the memory effect [1, 20], our approach is more robust. A single neural network can be trained to work for many different types of walls. It also does not require the object to be sparse, and there is no limit of field of view either. The simplicity and robustness of our method is enabled by direct recognition, which bypasses the imaging process.

## Discussion

We have shown that high-accuracy object recognition can be achieved by illuminating an object with coherent light and by analyzing the resulting speckle pattern from the scattered light using a deep-learning algorithm. Our approach eliminates the need for forming a clear image in the recognition process, and performs well even in situations where imaging is difficult or impossible. The resulting system may be highly desirable in mobile or autonomous systems, since it is highly data- and energy-efficient as compared to conventional imaging-based recognition systems. Our results also indicate that the theoretical limits for recognition are very different from the theoretical limits of imaging.

Nevertheless, we anticipate various issues in practice that could potentially affect performance of our system. As can be seen from the results without direct line-of-sight, the mechanical stability of the experimental components, such as the walls, could affect the recognition accuracy. The stability of speckle pattern and performance of the system could potentially be affected by air turbulence when applied to long-distance object recognition. Practical issues of the performance of our system such as these could be improved by employing more advanced neural network structures and deep learning method to increase the ability for the model to generalize to unseen situations and increase the robustness of the system. Integrating multiple frames during data collection could also be used to increase the SNR and the recognition accuracy.

Recent progress in deep learning has enabled record performance for analyzing complex scene images, which are often taken under good conditions. However, when used by autonomous systems in a degraded environment, visual object recognition can be challenging due to the fundamental limit of imaging optics. Removing the imaging step in visual recognition can open exciting prospects in challenging scenarios such as recognition without focusing, beyond diffraction limit, or without direct line-of-sight. Without taking, storing, and processing high-definition images, the imaging-free method can be highly data- and energy-efficient, which is highly desirable in mobile applications. We demonstrated that unlike incoherent light, coherent light directly carries scene information through its interference patterns. These complex patterns are often meaningless to humans but carry essential identity information to be recognized by deep-learning algorithms. Coherent illumination points to an exciting direction for direct visual recognition to be used in future autonomous systems.

# Acknowledgments

Z.Y. acknowledges the financial support from WARF.

# Author Contributions

Y.T., X.L., and K.X.W. performed the simulation and experiment. Y.T. and X.L. wrote the manuscript, with input from all authors. K.X.W. and Z.Y. supervised the project.

# References

1. Katz, O., Heidmann, P., Fink, M. & Gigan, S. Non-invasive single-shot imaging through scattering layers and around corners via speckle correlations. *Nat. Photonics* **8,** 784-790 (2014).
2. Katz, O., Small, E., Guan, Y. & Silberberg, Y. Noninvasive nonlinear focusing and imaging through strongly scattering turbid layers. *Optica* **1,** 170-174 (2014).
3. Katz, O., Small, E. & Silberberg, Y. Looking around corners and through thin turbid layers in real time with scattered incoherent light. *Nat. Photonics* **6,** 549-553 (2012).
4. Bertolotti, J. *et al.* Non-invasive imaging through opaque scattering layers. *Nature* **491,** 232-234 (2012).
5. Sinha, A., Lee, J., Li, S. & Barbastathis, G. Lensless computational imaging through deep

learning." *Optica* **4,** 1117-1125 (2017).

6. Horisaki, R., Takagi, R. & Tanida, J. Learning-based imaging through scattering media. *Opt. Express* **24,** 13738-13743 (2016).

7. Lyu, M., Wang, H., Li, G. & Situ, G. Exploit imaging through opaque wall via deep learning. Preprint at https://arxiv.org/abs/1708.07881 (2017).

8. Li, L. *et al.* Imaging through scattering layers exceeding memory effect range with spatial-correlation-achieved point-spread-function. *Opt. Lett.* **43,** 1670-1673 (2018).

9. Satat, G., Heshmat, B., Raviv, D. & Raskar, R. All photons imaging through volumetric scattering. *Sci. Rep.* **6,** 33946 (2016).

10. Guo, C., Liu, J., Wu, T., Zhu, L. & Shao, X. Tracking moving targets behind a scattering medium via speckle correlation. *Appl. Opt.* **57,** 905-913 (2018).

11. Kim, G. & Menon, R. Computational imaging enables a "see-through" lens-less camera. *Opt. Express* **26,** 22826-22836 (2018).

12. Cruz, J. M. D., Pastirk, I., Comstock, M., Lozovoy, V. V. & Dantus, M. Use of coherent control methods through scattering biological tissue to achieve functional imaging. *Proc. Natl. Acad. Sci.* **101,** 16996-17001 (2004).

13. Yoo, K. M., Xing, Q. & Alfano, R. R. Imaging objects hidden in highly scattering media using femtosecond second-harmonic-generation cross-correlation time gating. *Opt. Lett.* **16,** 1019-1021 (1991).

14. Borcea, L., Papanicolaou, G., Tsogka, C. & Berryman, J. Imaging and time reversal in random media. *Inverse Probl.* **18,** 1247 (2002).

15. Balaji, M. M., Viswanath, A., Rangarajan, P., MacFarlane, D. & Christensen, M. P. Resolving Non Line-of-Sight (NLoS) motion using Speckle. *Computational Optical Sensing and Imaging*, CM2E-2. Optical Society of America (2018).

16. Borhani, N., Kakkava, E., Moser, C. & Psaltis, D. Learning to see through multimode fibers. *Optica* **5**, 960-966 (2018).

17. Wang, P. & Di, J. Deep Learning-Based Object Classification through Multimode Fiber via a CNN-Architecture Speckle Net. *Appl. Opt.* 57, no. 28, 8258-63 (2018).

18. Horisaki, R., Takagi, R. & Tanida, J. Learning-Based Single-Shot Superresolution in Diffractive Imaging. *Appl. Opt.* 56, no. 32, 8896-8901 (2017).

19. Wu, Y., Rivenson, Y., Zhang, Y., Wei, Z., Günaydin, H., Lin, X. & Ozcan, A., Extended Depth-of-Field in Holographic Imaging Using Deep-Learning-Based Autofocusing and Phase Recovery. *Optica* **5**, 704-710 (2018).

20. Chang, J. & Wetzstein, G. Single-shot speckle correlation fluorescence microscopy in thick scattering tissue with image reconstruction priors., *J. Biophotonics.* 11(3) (2018).

21. Valent, E. & Silberberg, Y. Scatterer recognition via analysis of speckle patterns. *Optica* **5,** 204-207 (2018).

22. Kim, G., Kapetanovic, S., Palmer, R. & Menon, R. Lensless-camera based machine learning for image classification. Preprint at https://arxiv.org/abs/1709.00408 (2017).

23. Ando, T., Horisaki, R. & Tanida, J. Speckle-learning-based object recognition through scattering media. *Opt. Express* **23,** 33902-33910 (2015).

24. Takagi, R., Horisaki, R. & Tanida, J. Object recognition through a multi-mode fiber. *Opt. Rev.* **24,** 117-120 (2017).

25. Satat, G., Tancik, M., Gupta, O., Heshmat, B. & Raskar, R. Object classification through scattering media with deep learning on time resolved measurement. *Opt. Express* **25,** 17466-17479 (2017).

26. Goodman, J. W., Huntley Jr., W. H., Jackson, D. W. & Lehmann, M. Wavefront-reconstruction imaging through random media. *Appl. Phys. Lett.* **8,** 311-313 (1966).

27. Millane, R. P. Phase retrieval in crystallography and optics. *J. Opt. Soc. Am. A* **7,** 394-411 (1990).

28. Elser, V. Phase retrieval by iterated projections. *J. Opt. Soc. Am. A* **20,** 40-55 (2003).

29. Faulkner, H. M. L. & Rodenburg, J. M. Movable aperture lensless transmission microscopy: a novel phase retrieval algorithm. *Phys. Rev. Lett.* **93,** 023903 (2004).

30. Rodenburg, J. M. & Faulkner, H. M. L. A phase retrieval algorithm for shifting illumination. *Appl. Phys. Lett.* **85,** 4795-4797 (2004).

31. Zuo, J. M., Vartanyants, I., Gao, M., Zhang, R. & Nagahara, L. A. Atomic resolution

imaging of a carbon nanotube from diffraction intensities. *Science* **300,** 1419-1421 (2003).

32. Pfeiffer, F., Weitkamp, T., Bunk, O. & David, C. Phase retrieval and differential phase-contrast imaging with low-brilliance X-ray sources. *Nat. Phys.* **2,** 258-261 (2006)

33. Xu, X., Liu, H. & Wang L. V. Time-reversed ultrasonically encoded optical focusing into scattering media. *Nat. Photonics* **5,** 154-157 (2011).

34. Viswanath, A., Rangarajan, P., MacFarlane, D. & Christensen, M. P. Indirect Imaging Using Correlography. *Computational Optical Sensing and Imaging*, CM2E-3. Optical Society of America (2018).

35. Rangarajan, P. & Christensen, M. P. Imaging hidden objects by transforming scattering surfaces into computational holographic sensors. *Computational Optical Sensing and Imaging*, CTh4B-4. Optical Society of America (2016).

36. Zhao, J. *et al.* Bending-Independent Imaging through Glass-Air Disordered Fiber Based on Deep Learning. *Computational Optical Sensing and Imaging*, CW3B-6. Optical Society of America (2018).

37. Tyo, J. S., Rowe, M. P., Pugh, E. N. & Engheta, N. Target detection in optically scattering media by polarization-difference imaging. *Appl. Opt.* **35,** 1855-1870 (1996).

38. Velten, A. *et al.* Recovering three-dimensional shape around a corner using ultrafast time-of-flight imaging. *Nat. Commun.* **3,** 745 (2012).

39. O'Toole, M., Lindell, D. B. & Wetzstein, G. Confocal non-line-of-sight imaging based on the light-cone transform. *Nature* **555,** 338-341 (2018).

40. Psaltis, D. & Farhat N. Optical information processing based on an associative memory model of neural nets with thresholding and feedback. *Optics Letters* 10 (2), 98-100 (1985)

41. Chang, J., Sitzmann, V., Dun, X., Heidrich, W. & Wetzstein G. Hybrid optical-electronic convolutional neural networks with optimized diffractive optics for image classification. *Scientific reports* 8 (1), 12324 (2018).

**Figure 1** Information carried by coherent and incoherent illuminations. **a** Samples of objects to be recognized, taken from the MNIST database. **b** Object is illuminated by coherent light. The scattered light intensity distributions are shown at different locations. Neural networks are used to recognize the digits from these field intensity distributions, with a high recognition accuracy (>90%). **c** The same object is illuminated by incoherent light. It produces nearly uniform field distributions, leading to a recognition accuracy as low as a random guess (10%). **d** Schematic of the sampling schemes used by imaging and imaging-free systems.

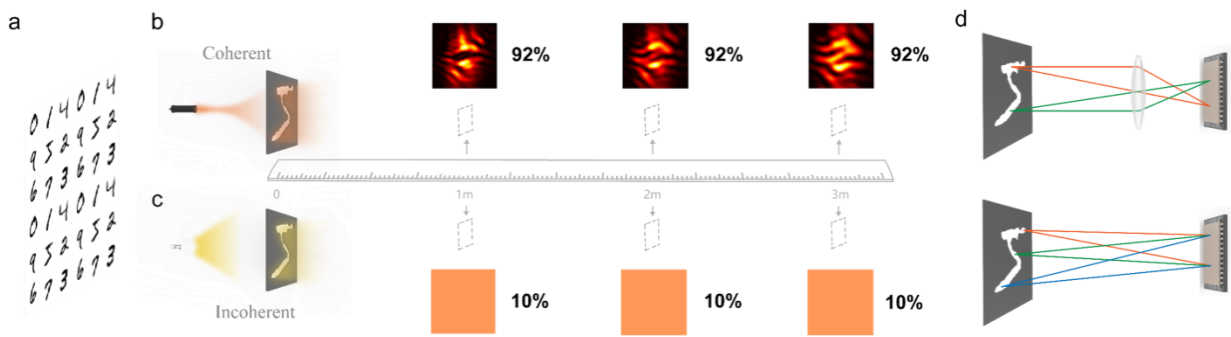

**Figure 2** Focusing-free recognition. **a, b** Schematic setup of experiments, illuminated by a white lamp with an imaging system (a) and by an expanded laser beam (b). A CMOS image sensor is used to measure the intensity pattern distribution of reflected light. For both coherent and incoherent illuminations, the object starts at 1 m away from the sensor and then moves away, all the way to 8 m. Scattered patterns of the object at different locations are recorded. **c, d** Intensity patterns of light taken by the CMOS image sensor under incoherent (c) and coherent (d) illumination. Patterns of different objects (1st row for object "2" and 2nd row for object "9") and same object at different locations (from 1 m at leftmost to 8 m at rightmost) are shown. **e, f** Neural network recognition accuracy at different object–sensor distances under coherent vs. incoherent illumination in simulation (e) and experiment (f). In the experiment, we use one network for all locations for both coherent and incoherent illuminations. The single network performs well for all distances with coherent illumination and fails for incoherent illumination. In the simulation, we train separate neural networks for each location for incoherent illumination to maximize its recognition rate.

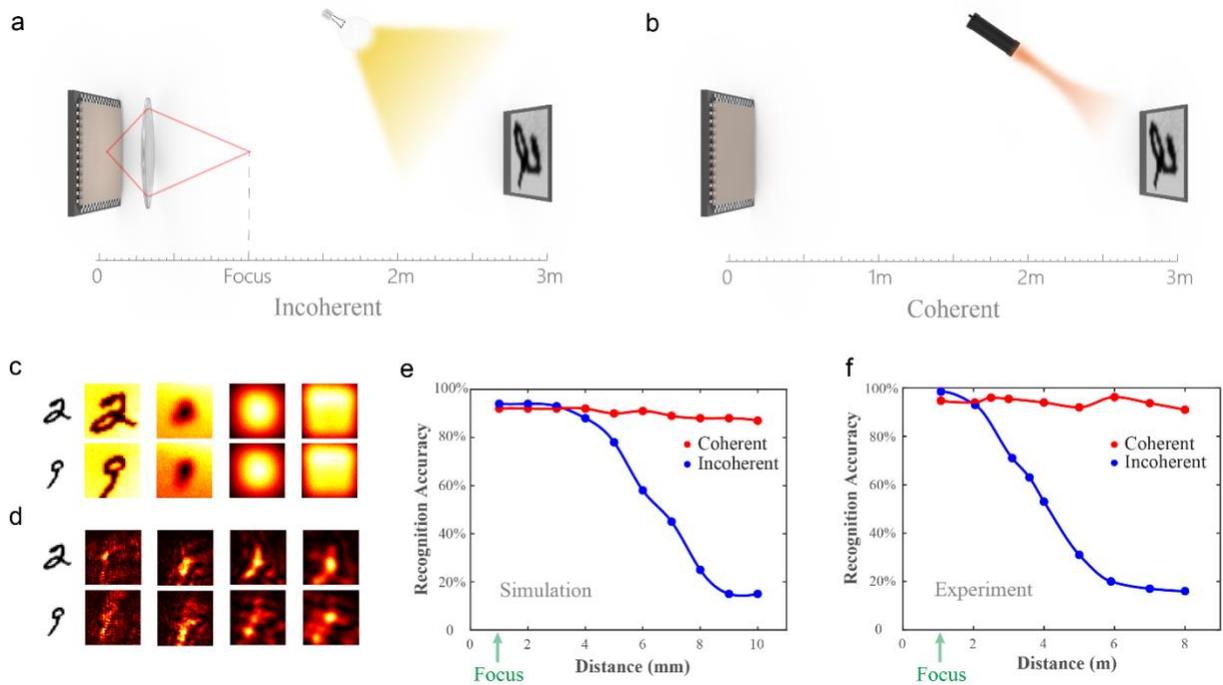

**Figure 3** Recognition beyond diffraction limit. **a, b** Schematics of setup. Object with shape of hand-written digits is illuminated by coherent (a) and incoherent (b) light. An aperture with diameter D is placed in front of the object at a distance of L. A CMOS image sensor is used to measure the intensity pattern distribution of transmitted light. With coherent illumination, the intensity distribution is measured directly. With incoherent illumination, a lens is placed after the aperture to focus the object onto the CMOS image sensor. For both coherent and incoherent illuminations, the size of the aperture is gradually reduced and intensity patterns of different aperture sizes are recorded. **c** Neural network recognition accuracy with different apertures under coherent vs. incoherent illumination in simulation. **d** Neural network recognition accuracy with different apertures under coherent illumination in the experiment.

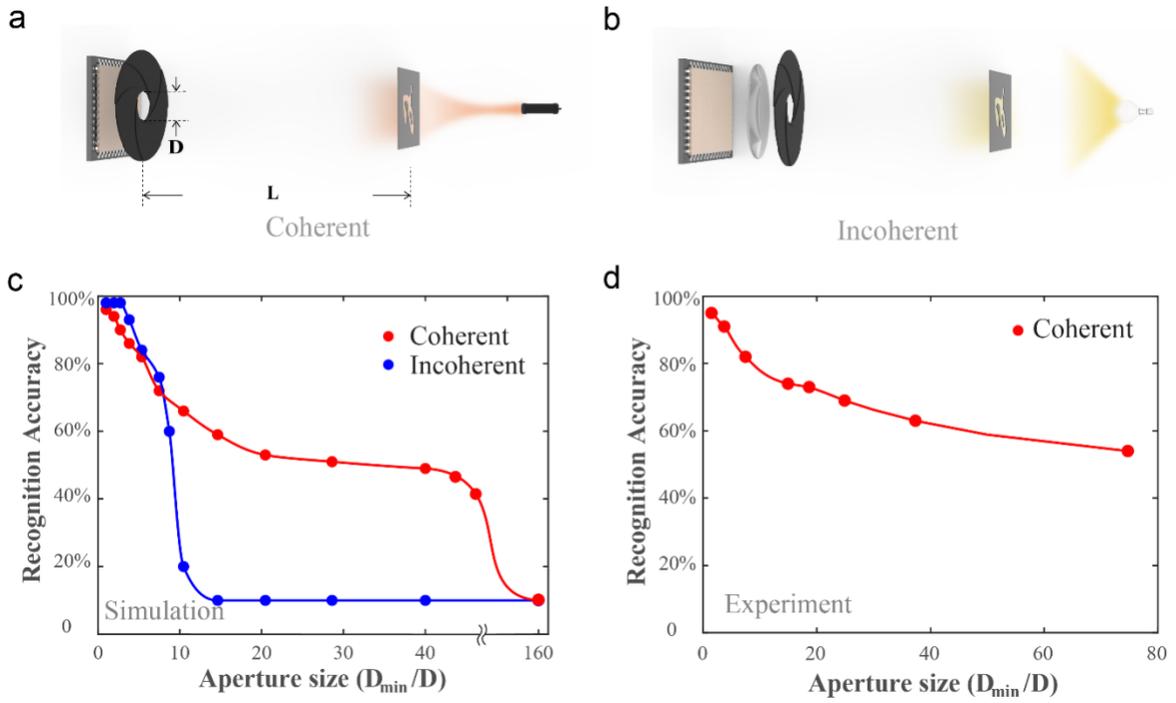

**Figure 4** Recognition without direct line-of-sight. **a** Schematic setup of the experiment. Direct line-of-sight between screen and image sensor is blocked. Image sensor captures light from LCD screen through scattering by a diffusive surface (wall). **b** Images of different hand-written digits captured on the CMOS image sensor. **c** Confusion Matrix of neural network classification results on test objects. Separate accuracies are shown for objects in different categories of hand-written digits. The overall recognition accuracy is greater than 80%.

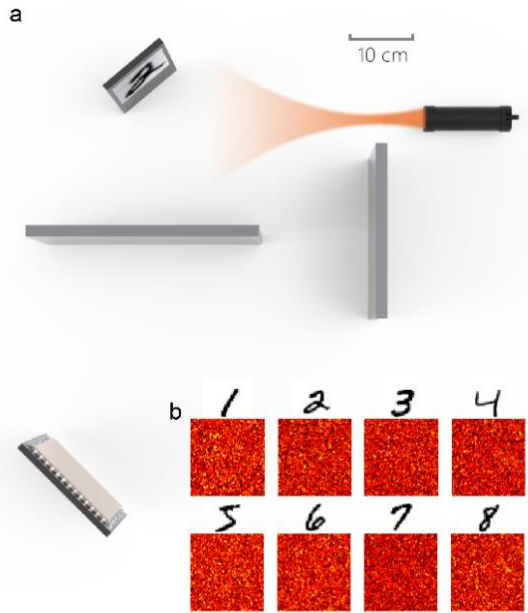 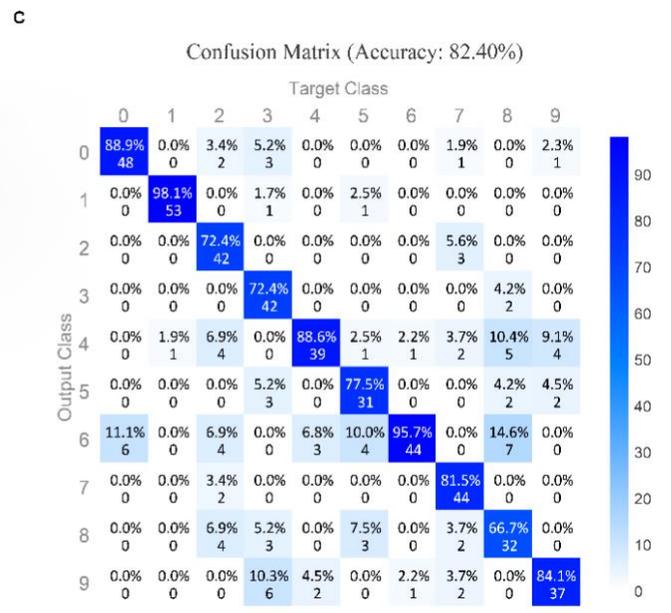